\documentclass[%
 reprint,
 amsmath,amssymb,
 aps,
]{revtex4-2}

\usepackage{graphicx}% Include figure files
\usepackage{dcolumn}% Align table columns on decimal point
\usepackage{bm}% bold math
\usepackage{xcolor}

\begin{document}

\preprint{APS/123-QED}

\title{Ultrafast Electron Dynamics in Magnetic Thin Films}
%\title{Ultra-fast Magnetisation Quench in Nickel Induced by Low Energy Structural Dynamics}
%\title{Terahertz-Induced Ultrafast Structural Dynamics in Ferromagnetic Nickel Film}% Force line breaks with \\

\author{Hovan Lee}
\email{hovan.lee@kcl.ac.uk}
\author{Cedric Weber}%
\affiliation{%
 King’s College London, Theory and Simulation of Condensed Matter, The Strand, WC2R 2LS London, UK
}%


\author{Manfred Fähnle}
\affiliation{71272 Renningen, Germany, Schönblickstraße 95, former member of the Max Planck Institute for Intelligent Systems, Stuttgart, Germany}
%\affiliation{
%\affiliation{
 %Forth institution, the second for Charlie Author
%}%

%\affiliation{King's College London, The Strand, WC2R 2LS London, UK}
\author{Mostafa Shalaby}
\email{most.shalaby@gmail.com}
\affiliation{Beijing Key Laboratory for Precision Optoelectronic Measurement Instrument
and Technology, School of Optics and Photonics, Beijing Institute of Technology, Beijing 100081, China}
\affiliation{Swiss Terahertz Research-Zurich, Technopark, 8005 Zurich, Switzerland and Park Innovaare, 5234
Villigen, Switzerland}


\date{\today}% It is always \today, today,

\begin{abstract}
In past decades, ultrafast spin dynamics in magnetic systems have been associated with heat deposition from high energy laser pulses, limiting the selective access to spin order. Here we use a long wavelength terahertz (THz) pump – optical probe setup to measure structural features in the ultrafast time scale. We find that complete demagnetisation is possible with $<6 THz$ pulses. This occurs concurrently with longitudinal acoustic phonons and an electronic response, followed by the magnetic response. The required fluence for full demagnetisation is low, ruling out the necessity of a high power light source.

%In past decades, ultrafast electron dynamics in thin films have been associated with entangled processes of the charge, spin and lattice dynamics, limiting the ability to study the degrees of freedom independently. Here we use a low energy terahertz (THz) pump – optical probe setup under different magnetic field conditions to measure structural features of nickel thin films in the ultrafast time scale.
%We find complete demagnetisation with $<6 THz$ pulses. This is mediated with longitudinal acoustic phonons and an electronic response, followed by the magnetic response. The required fluence for full demagnetisation is low, ruling out the necessity of a high power light source.

%\begin{description}
%\end{description}
\end{abstract}

\maketitle

%\tableofcontents

\section{\label{sec:level1}Introduction}

The study of ultrafast magnetism is crucial in advancing our understanding of magnetic systems, as well as developing ultrafast memory devices. The difficulty of this subject lies in spin dynamics, and its role as a part of a larger picture of the barely understood coupled structural dynamics. 

As an example, when a ferromagnetic film is excited by a femtosecond (fs) laser pulse, partial demagnetisation of the material occurs within $\sim 100 fs$, followed by a remagnetisation to the original state on a longer time scale. This was first observed in 1996 for optical laser pulses \cite{manfred1}, and later also for THz laser pulses \cite{mostafa_ni,thz,thz2,referee1}. %cite3
While this phenomenon has led to intense investigations in the field by both experimentalists and theorists, it has now become evident that this process is mediated by a complex puzzle of strongly entangled pieces \cite{manfred2}. To elaborate on this, it is essential to understand the physical mechanisms that can influence the spin angular momenta of the electrons responsible for magnetisation.

\subsection{\label{sec:level2}Separation of Underlying Mechanisms}

In order to simplify the almost unfeasible task of tackling coupled structural dynamics, we examine the phenomena responsible and their respective timescales. This allows us to decouple mechanisms which have infinitesimal effects on each other.

% This file may be formatted in either the \texttt{preprint} or
% \texttt{reprint} style. \texttt{reprint} format mimics final journal output. 
% Either format may be used for submission purposes. \texttt{letter} sized paper should
% be used when submitting to APS journals.
As Born and Oppenheimer 
%/ M. Born and J. R. Oppenheimer , Ann. Phys. 389, 457 (1927)/ 
pointed out, a separation of atomic nucleus dynamics and the surrounding electrons can be made. This is because electrons will adiabatically follow any comparatively slow changes of the lattice. That is, typical dynamics of the phonons is on the picosecond timescale, where as electron dynamics in metals can occur in as short a time span as the attosecond timescale.
% https://www.sciencedirect.com/science/article/pii/S0368204820300116

Similarly, a separation of electron charge and spin degrees of freedom is reasonable, due to the slower time scales that are observed for the dynamics of the spins, $\sim 100 fs$, than that for the electron scattering, $\sim 10 fs$. 
%Time scales can be quite different for each of these processes. 

\begin{figure*}
    \centering
    \includegraphics[width=0.9\linewidth]{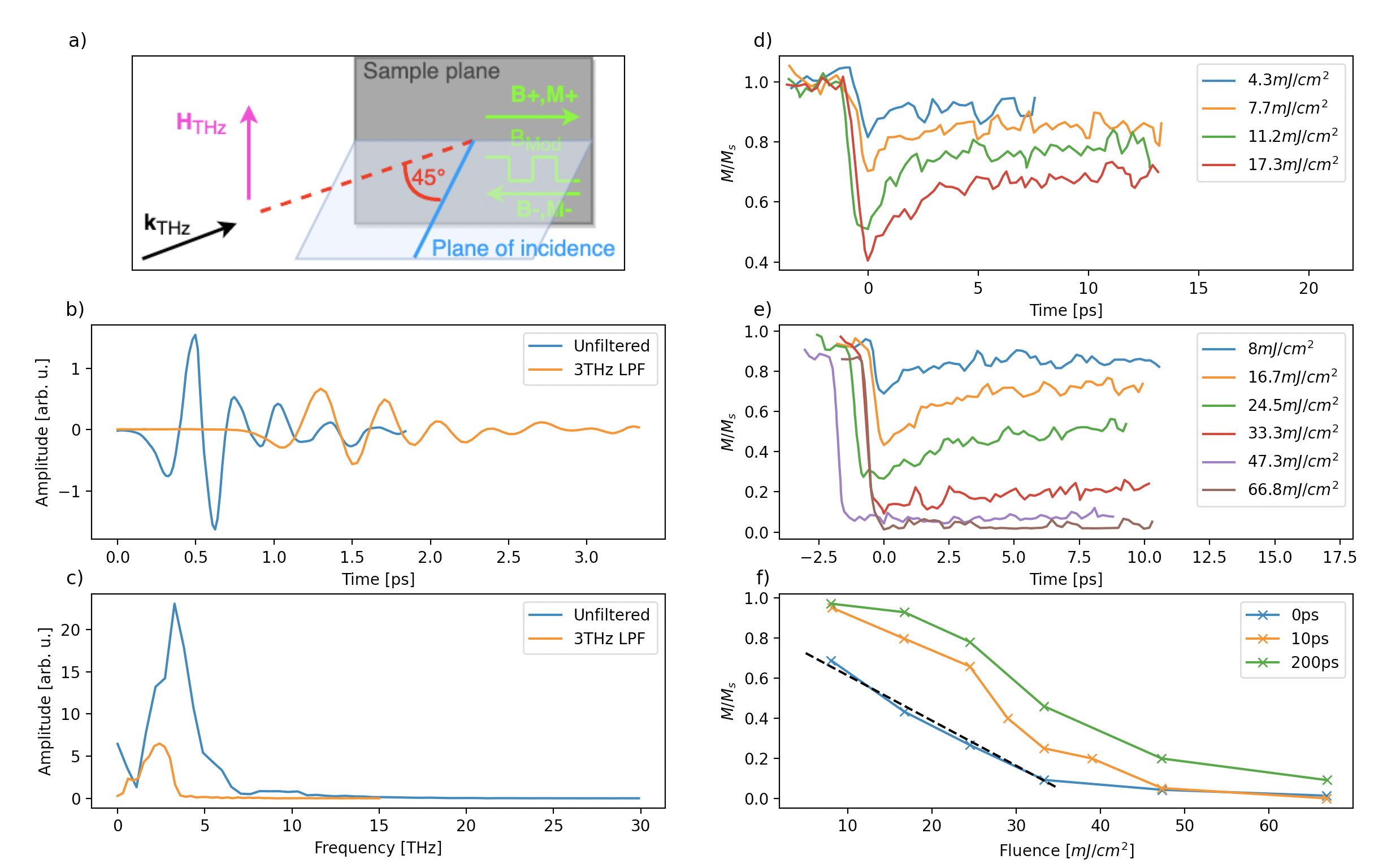}
    \caption{(a) The experimental setup for THz pump and optical MOKE probe. (b) The triggering unfiltered THz pulse. (c) Fourier transform of the unfiltered THz pulse (d) Terahertz-induced demagnetisation taken with 3 THz low pass filter and THz excitation levels of 4.3, 7.7, 11.2 \& 17.3 $mJ/cm^2$. (e) The demagnetisation measurements taken with 6 THz low pass filter for 8, 16.7, 24.5, 29, 33.3, 39, 47.3 \& 66.8 $mJ/cm^2$. (f) The fluence-dependent demagnetisation taken with 6 THz low pass filter for delay of 0,10 and 200 ps, with a straight line as a guide for the eye to showcase the near linear response.}
    \label{f1}
\end{figure*}

\subsection{\label{sec:level2}Pathways of Magnetisation Dynamics}

%Moreover, the spin demagnetisation might involve electron and lattice (phonon) degrees of freedom.
The dynamics of magnetism in a stimulated material could be affected by any number of combinations of the mechanisms described above.
In particular, there may be different channels of light-matter interactions, here we elaborate on the two extremes: direct coupling of the laser pulses to electronic spins, and intermediate charge and/or phonon excitations which induce magnetisation dynamics \cite{electron-phonon,optical,manfred1,manfred6,electron-phonon2,electron-phonon3,electron-phonon4}.

In the direct scenario, the local electro-magnetic field of the stimulus directly couples with the electronic spin angular momenta \cite{manfred5,manfred6},
%This is the case for THz laser pulses of small field amplitudes
such that there is minimal heating of the electronic structure.

In the indirect channel, the laser photons do not primarily change the magnetisation, but instead photon energy is transferred to the system in the form of an increased electronic and/or atomic temperature. This process can excite electron-hole pairs, lattice phonons and magnons, which will non-selectively spin-flip scatter with the electrons responsible for magnetisation, modifying their angular momenta \cite{spin-flip,spin-flip2,spin-flip3,magnon}.

Demagnetisation effects in past reports with optical laser pulses are mediated through this indirect channel. %called superdiffusion \cite{manfred3}.
This results in a non-equilibrium state, where the electronic temperature is raised above the phononic temperature, and an imbalance exists between chemical potentials of electrons with differing spins. This difference is the driving force for ultrafast demagnetisation \cite{manfred4,chem_pot}.
%It is also the case, that the temperature of the electrons is higher than the temperature of the phonons.
The system then evolves through the above discussed spin-flip scatterings, leading to changes in the orientations of atomic magnetic moments. As a result of this, a temporary demagnetisation is observed, followed by remagnetisation to the original state through the balance of the chemical potentials, and the thermalisation of the electrons and phonons.
%there may be a direct change of the magnetisation induced by the direct interaction of the local magnetic moments with the wave. This is the case after THz laser pulses with small field amplitudes. 
Furthermore, in samples with metallic substrates there is also a contribution of a superdiffusion process \cite{manfred3}. In this process, the excited mobile spin carriers are transferred onto a conducting substrate.
%%%%%%%%%%%%%%%%%%%% read M. Battiato, K. Carva, and P. M. Oppeneer, Superdiffusive spin transport as a mechanism of ultrafast demagnetization, Phys. Rev. Lett. 105, 027203 (2010).
%%%%%%%%%%%%%%%%%%%% and write this paragraph

Lastly, the total effective magnetic field exerts a Zeeman torque onto the atomic magnetic moments. This minute contribution on the magnetisation $M(t)$ is a coherent oscillation in time, following the action of the electro-magnetic wave. However, excited electrons with the above discussed spin flip scattering alter $M(t)$ incoherently in time, leading to fluctuations of the phase; dephasing the system. Therefore, the precessional motion is not typically observable after optical laser pulse excitations. 
\textcolor{black}{Although the effects of Zeeman torque has been disproven to influence electronic spins to the point of demagnetisation in the Mott insulator NiO \cite{kampfrath}, this effect has not been shown in metals.}

%%%%%%%%%%%%%%%%%%%% add thz pulse from mostafa

\subsection{\label{sec:level2}Unto Terahertz Laser Pulses}

Although optical pulse / thin film demagnetisation have been intensively examined, it has been shown that the majority of these studies rely on indirect and indiscriminate spin excitations \cite{temp,temp2,ey}. To increase the coherency and control of the interaction, experiments have been preformed with THz laser stimuli. The THz field cycle oscillates on a similar timescale as the natural speed of electronic spin motions, as opposed to optical pulse stimuli, which oscillate at a much faster timescale.

\begin{figure*}
    \centering
    \includegraphics[width=0.8\linewidth]{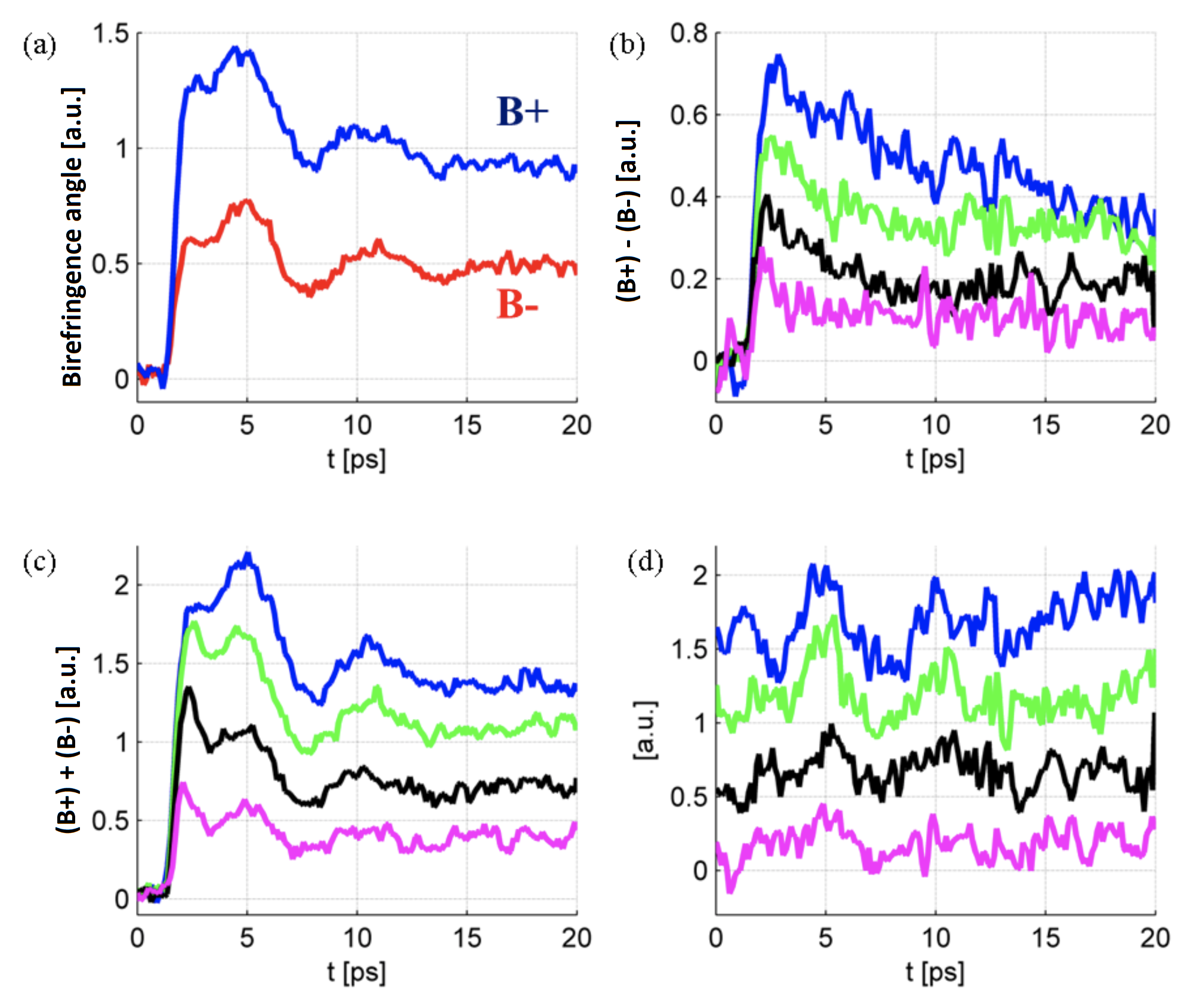}
    \caption{(a) The THz-induced birefringence in Ni using an external mechanical chopper (CM) under and external magnetic bias B+ (CM+) \& B-(CM-) with a fluence of $190 mJ/cm^2$. (b) The extracted magnetic response (demagnetisation) obtained from CM+ - CM-. Blue, green, black, and magenta refer to the fluences of 47, 85, 123, \& $190 mJ/cm^2$. (c) the total electronic and structural response is obtained using (CM+) + (CM-) which contains both thermal excitation and acoustic phonons. The latter is independently shown in (d).}
    \label{f2}
\end{figure*}

At low field amplitudes, THz lasers are therefore expected to directly and coherently couple with electronic spin dynamics. This interaction leads to a precessional motion of $M(t)$ due to the Zeeman torque, allowing for selective control of the magnetic phase.
Moreover, THz photon energies are three orders of magnitude smaller than optical photon energies, inducing significantly less heating at low field amplitudes, and reducing the possibility for spin flip scattering between electrons and excited particles. This direct coupling of THz and demagnetisation was reported for ferromagnetic cobalt films \cite{cobalt}, where a coherent, phase locked demagnetisation response was observed under excitations from THz pulses.

%Although optical pulses and their relation with demagnetisation have been intensively studied, probing and controlling spin dynamics with THz pulses remains challenging. For THz laser pulses, a distinction applies between respectively small and high field amplitudes. For the former, there is a coherent interaction between the magnetisation and effective magnetic field of the pulse $B_{eff}$, which oscillates in time due to the action of the THz laser pulse. In this case, there is a direct correlation between the temporal magnetisation dynamics and the temporal oscillation of the laser field /3/, in contrast to the lack of such a temporal correlation after optical laser pulses. This coherent interaction leads to a precessional motion of the magnetisation due to the Zeeman torque. At low field excitations, this precessional motion becomes observable because of the lack of a strong heating of the sample in THz (the THz photon energies are about three orders of magnitude smaller than the optical photon energies).

With the recent advances of THz source technology, THz laser pulses with field amplitudes of several Teslas are now accessible. This opens new venues for exploring complete precessional reversal of $M(t)$. From a technological perspective, obtaining such a reversal would be a milestone, as the time scales involved would be significantly shorter than the Larmor precession obtained in present-day technology. %constant magnetic field.
Clearing a potential pathway to increase processor speeds.  %%%%%%%%%%%%%%%%%%%%%% NOT SURE IF THIS PARAGRAPH FITS HERE %%%%%%%%%%%%%%%%%%%%%%%%%

However, high field amplitudes come at the cost of a strong heating of the sample, and a loss of the coherent interaction between $M(t)$ and the THz laser field. In this regime, incoherent electronic excitations with subsequent spin-flip scatterings induce a dephasing of M(t) without precessional motion. This is similar to the processes observed after optical pulses. Finally, an additional challenge at large field comes in the form of permanent modifications of film properties, and possible damage observable with scanning electron microscopy.

Here, we approach the ultrafast magnetisation dynamics from the photon energy perspective, where we compare excitation dynamics from high (optical) to low (THz) photon energies. We experimentally study the THz induced ultrafast electronic and structural dynamics in nickel thin films.

%An optically induced demagnetisation was reported from [xx et al], where xx mJ/cm2 was sufficient for complete demagnetisation in a 15 nm-thick Ni film. 

In the THz frequency range, partial demagnetisation in Ni has been previously reported, followed by permanent demagnetisation \cite{mostafa_ni}. Here, we carefully design the experiment to accurately map the THz-induced demagnetisation. We report that sufficiently intense THz pulses lead to full demagnetisation on the ultrafast time scale without sample damage. The process is combined with strong electronic excitation and generation of acoustic phonons.

\textcolor{black}{With the knowledge required to understand the principle mechanisms governing the effects of THz laser pulses on ferromagnetic materials, we move on to the results of our experiment.}

\section{\label{sec:level1}Results}

For our experiment, a $15nm$ sputtered Ni thin film on a high resistivity Si substrate was chosen as the sample. This is due to the many extensive studies performed upon the ultrafast magnetism of the material, first reported by Beaurepaire et al in 1996.

The setup is a tightly focused terahertz bullet scheme where the THz is focused to the diffraction limit to reach the maximum possible intensity. The THz is generated by optical rectification of near infrared pulses (1550 nm central frequency, 100 Hz, 50 fs, 3.5 mJ, Light Conversion OPA system) and an organic crystal DAST (Swiss Terahertz GmbH), 350 um thick. The beam is expanded then focused on the sample (three off axis parabolic mirrors scheme). The probe beam is 800 nm centered, 70 fs, 100 Hz, and collinear with the THz pump. The sample was placed to allow for an incident angle of 45. Both probe and pump were linearly polarized, and the full setup is illustrated in Fig.\ref{f1}.a. The THz-induced magnetisation dynamics lead to birefringence on a collinear $800 nm$ probe. The temporal and spectral contents of the THz pulse are shown in Fig.\ref{f1}.b-c. All measurements of the ratio between the magnetisation and its saturation value $M(t)/M_s$, as shown in Fig.\ref{f1}, were performed by modulating the external magnetic field at $40 mT$ and $25 Hz$. This frequency was used as a reference for our acquisition system to eliminate any non-magnetic contribution to the measured signal. This exact experimental setup was characterised in depth in \cite{mokekerr}.

%As a sample, we chose Ni thin film (15 nm on high resistivity silicon substrate). Ni has been well studied in the field of ultrafast magnetism since the first reported in Bauerpeier et al in 1996.
%The experimental setup involves THz generated through optical rectification of near infrared fs pulses in organic crystal DSTMS (details of the source are specified in \cite{thz_bullet}) and a Magneto-Optical-Kerr-Effect (MOKE) probe of $30\mu m$ diameter as shown in Fig.\ref{f1}.a. The THz-induced magnetisation dynamics lead to birefringence on a collinear $800 nm$ probe. The temporal and spectral contents of the THz pulse are shown in Fig.\ref{f1}.b-c. All measurements of the ratio between the magnetisation and its saturation value $M(t)/M_s$, as shown in Fig.\ref{f1}, were performed by modulating the external magnetic field at $40 mT$ and $25 Hz$. This frequency was used as a reference for our acquisition system to eliminate any non-magnetic contribution to the measured signal. This exact experimental setup was characterised in depth in \cite{mokekerr}.

%\subsection{\label{sec:level2}Determination of Demagnetisation Mechanism}

The THz source contains significant spectral components up towards $\sim18THz$, these components were eliminated through a set of low pass filters with cut-off frequencies at $3$, $6$ and $9 THz$. This results in a set of pulses that are well defined in spectral content, we  expect these pulses to contain several oscillation pulses in the time domain. 
To further characterise the pulses, we refer to a previous work on the same THz source: \cite{diamond}. In the cited text, a one-to-one correspondence between the pulse fluence and the peak electric field strength was found through the Kerr effect in diamond, allowing us to infer the fluence of a filtered pulse by measuring the pulse duration and spot size (measured previously in \cite{thz_bullet}).

Through the $3THz$ filter, the maximum achievable THz fluence on the sample is $17.3 mJ/cm^2$, leading to an instantaneous demagnetisation (dM/Ms) of 60\% as shown in Fig.\ref{f1}.c.
Extending the spectrum with the $6 THz$ filter, presented in Fig.\ref{f1}.d, allows for much higher peak fluence up towards $66.8 mJ/cm^2$. This offers the possibility of complete demagnetisation without sample damage at $47 mJ/cm^2$, as alluded to in Fig.\ref{f1}.e, where magnetisation is shown to recover.
\textcolor{black}{Measurements of the reflected pulse show that around 90\% of the pulse fluence was reflected. This suggests that roughly 10\% of the incident fluence is absorbed. Therefore, this corresponds to an absorbed fluence of $\sim 4.7 mj/cm^2$}.
%Examining the fluence-dependent demagnetisation (Fig. 1d & 1e on the short and long temporal delays) shows the possibility of complete demagnetisation for the  $<6 THz$ excitation at fluence of $47 mJ/cm^2$ without sample damage.

At low fluence, the extent of demagnetisation increases with the excitation fluence almost linearly, suggesting a linear THz absorption mechanism in Fig.\ref{f1}.f.
%However, the magnetic response starts to deviate from linearity as the excitation fluence increases, as shown in Fig. 1f. We attribute this to nonlinear absorption mechanisms.
%However, nonlinear absorption is observed at higher excitation fluence, this is indicated by the deviation of magnetic response from linearity in Fig.\ref{f1}.f.
%start here

Here, we perform a rough calculation to compare the sample temperature at various fluences with the Curie temperature of Ni. The heat equation of the system is: $Q=m*c*dt$, where m is the mass of the object, c is the heat capacity, and dt is the induced temperature change. Bulk nickel has a heat capacity of $0.44 J/gK$ and a density of $8.908x10^6 g/m^3$. In our experiments, the sample thickness is $15 nm$ and we assume that about 10\% of the incident THz pulse is absorbed by the sample. Using a low pass filter of 6 THz we observe partial demagnetisation at a fluence of $16.7 mJ/cm^2$, corresponding to a temperature increase of ~280k. Complete demagnetisation was observed at $47.3 mJ/cm^2$, corresponding to a ~800k temperature increase. Taking into account that the experiment was performed at room temperature, this gives us the temperature range of ~500K to ~1070k, which is comparable to the Curie temperature of Ni at 627K.
%\textcolor{red}{This identifies that the the demagnetisation observed at very large fluence is due to the existence of hot electrons at the Fermi level, similar to previously reported for optical laser sources, but has not been proven previously for THz sources.}
This identifies that the the demagnetisation observed at very large fluence is due to the existence of hot electrons at the Fermi level (whereby electrons are excited by the intense sub ps heating out of the Fermi level. These electrons are prevented from thermalising with the Fermi sea due to the non equilibrium nature of the system, and therefore dominates the subsequent dynamics \cite{hot_e1,hot_e2}). This is similar to previously reported for optical laser sources, but has not been proven previously for THz sources.

Previous results on a similar sample \cite{mostafa_ni} showed a maximum demagnetisation of 58\% at $89mJ/cm^2$ before the sample becomes permanently damaged. This is in contradiction to the results we present here, where higher demagnetisation was achieved at lower THz fluence without noticeable damage.
%We explain that by the (relatively large) probe undersampling of the small THz spot reached with the spectral contents between 6 THz (this report) and previous report (18 THz). In other words, full demagnetisation in [xx] likely occurred without the probe being able to resolve it until the demagnetisation fluence was reached. Limiting the spectral contents to sub-6 THz in our experiment prevented the minimum THz spot size from being too small to be fully resolved by the optical probe.
We offer an explanation to this discrepancy: The MOKE probe in the previous report likely lacked the spatial resolution to pinpoint the demagnetised area exactly.
Hence both the areas of demagnetisation and the areas that are unaffected contributed to the measurement, giving a lower demagnetisation percentage. In this paper we limit the spectral contents to $<6THz$, this increases the diffraction-limited excitation spot size. This allows the area of demagnetisation to be fully resolved, and therefore achieving higher measurable demagnetisation at lower fluence.
%This can be accounted for by the relatively large dimension of the probe used in previous experiments, which was likely undersampling the small thz spot.

\begin{figure}
    \centering
    \includegraphics[width=\linewidth]{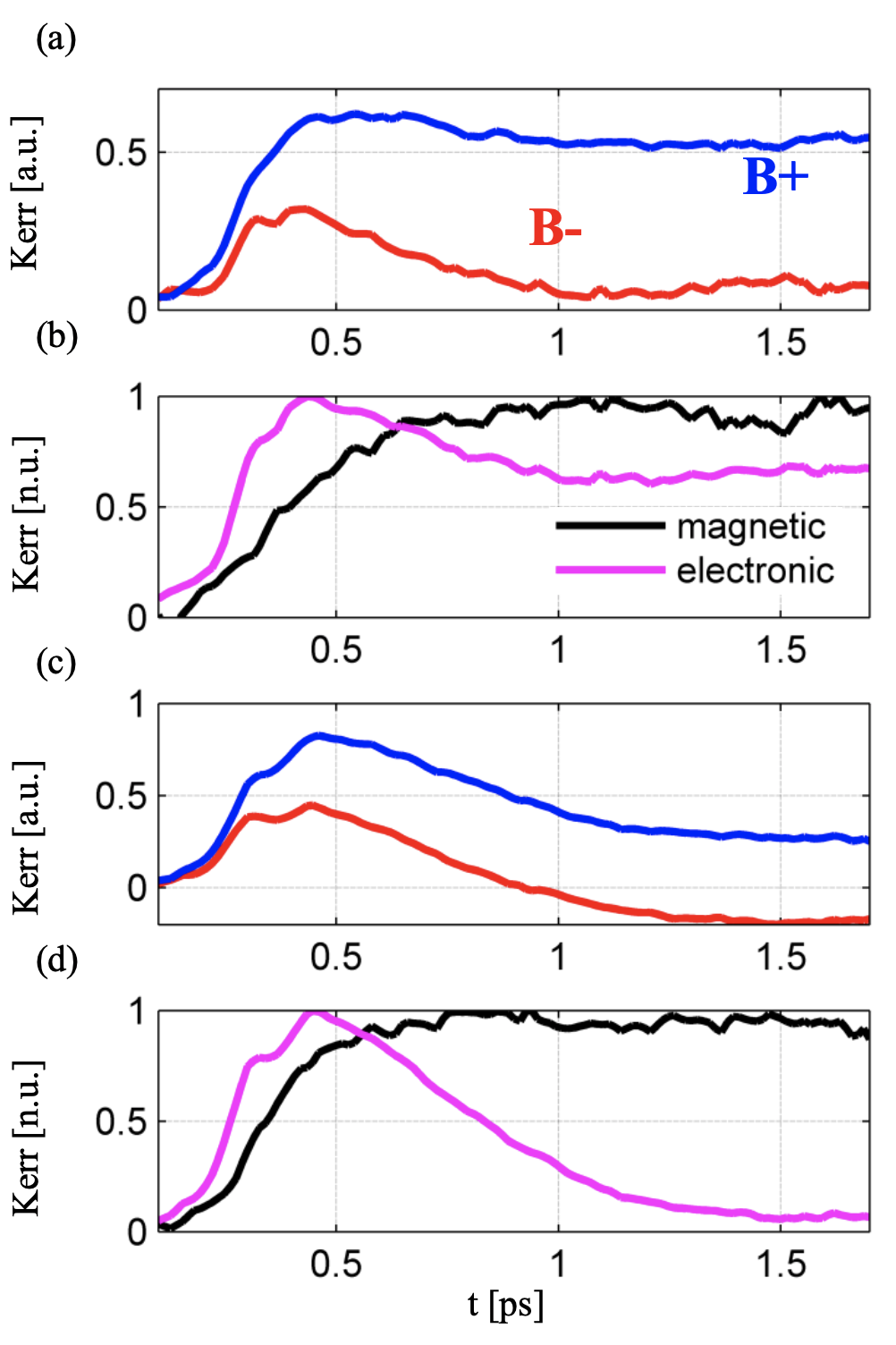}
    \caption{(a) The measured Kerr signal (CM+ and CM-) under a $9 THz$ filtered fluence of $48 mJ/cm^2$. (b) The extracted magnetic (CM+ - CM-) and electronic (CM+ + CM-) responses. (c) \& (d) are the corresponding measurements to (a) \& (b) under a fluence of $85 mJ/cm^2$.}
    \label{f3}
\end{figure}

We use this quantitative result to address the main question of the role of photons in demagnetisation. Specifically, whether demagnetisation was achieved through the direct or indirect channel discussed above. 
% There are two possible physical mechanisms responsible for the modifications of the spin angular momenta of the electrons of the sample: Firstly, there may be a direct change of the magnetisation by its interaction with the exciting wave, a coherent process generated by the Zeeman torque exerted by the total effective field $B_eff$ (which oscillates in time due to the action of the optical electromagnetic wave) on the magnetisation $M(t)$. Second, the laser photons indirectly change the magnetisation by exciting electron-hole pairs, with subsequent modifications of their spin-angular momenta. 
Conventionally, it was assumed that the direct mechanism was associated with excitations of low frequency electro-magnetic radiations, such as microwave radiation.
In contrast, high frequency optical pulse excitations have been shown to demagnetise materials through the indirect mechanism dominantly.
% In contrast, the indirect mechanism is the dominant one after optical laser pulse excitation with high frequencies.
However, it is not impossible for both mechanisms to coexist.
%The transition between the two is however not abrupt but instead a crossover.

It is generally considered that the higher the excitation pulse frequency, the higher the excitation field amplitude is needed in order to induce changes in the magnetisation.
%A good approximation is that the required field amplitude to induce a change in the magnetisation is an inverse function of the exciting wave frequency.
Therefore, to achieve magnetic switching in the THz regime, it would require practically unrealistic pulse intensities ($>10$ Tesla).
In this limit, undesired structural changes would inevitably take place. Our experiments show that THz excitations are only capable of inducing small amplitude precessions, as is confirmed in Fig.\ref{f1} where no precessions are observed in the measured signal. Therefore, our results point towards the indirect mechanism as the most likely hypothesis. 

A lingering question remains: Do individual photon energies play a significant role towards the measured demagnetisation effect.
%However, this raises the question on whether the photon energy plays a significant role towards the measured demagnetisation effect.
Indeed, this indirect mechanism critically depends on the presence of a high integrated photon energy to raise electron temperatures above phonon temperatures.
%Indeed, this indirect mechanism critically depends on having energetically enough photons to heat electrons above the phonon temperatures.
Comparing the photon energies from the optical ($eV$ energy scale) range to the THz ($meV$ energy scale) range, we conclude that this effect is not directly possible without exponentially increasing the number of THz photons. 
%Extending this scenario from the optical ($eV$ energy scale) to the THz ($meV$ energy scale) range is thus not directly possible in a linear fashion without exponentially increasing the number of photons. 

In contrast to this claim, the experimental results do not reveal a relationship between photon energy and demagnetisation. First, there is no high fluence threshold for demagnetisation; A nearly linear relationship between fluence and demagnetisation is observed in our experiments (Fig.\ref{f1}.f).
Secondly, with comparisons between our results and a previous report on a similar sample under optical excitation, the absorbed fluence needed for demagnetisation from optical and terahertz pulses are of the same order of magnitude. The difference in energy between individual optical and THz photons are three orders of magnitude apart, and thus confirms a negligible dependence of demagnetisation on individual excitation photon energy, ruling out any significant direct interaction of the THz electric field frequency on demagnetisation in our experiments. With evidence suggesting that only the net fluence absorbed influences the demagnetisation in the sample.

In light of these results, we extended our search of the mechanisms behind demagnetisation by including charge and lattice degrees of freedom. Ultrafast excitation of thin films leads to rapid heating and excitation of electrons to higher non-equilibrium states. This absorbed energy is eventually transferred to the lattice through electron-phonon interactions, leading to spatio-temporal strain pulses and coherent acoustic phonons which reverberate inside the film. This model has been used to explain ultrafast optical generation of lattice strain waves of coherent acoustic phonons, which can manipulate and coherently control the magnetisation orientation in ferromagnetic films \cite{electron-phonon,electron-phonon2,electron-phonon3,electron-phonon4}. Thereby justifying our investigation into these additional dynamics.

To tackle this obstacle, we decoupled the intricately intertwined effects of charge, spin and phononic dynamics, without knowing their specific relationships with one another, by exploiting the timescale differences between them. This was achieved by modulating the THz pump excitation and repeating the experiment in Fig.\ref{f1} with static magnetic fields of $B_+=+40mT$ and $B_-=-40mT$, corresponding to biases which are parallel and antiparallel to the THz pulse magnetic field, thus maintaining a constant magnetic bias during our experiments.

The dynamics under the different biases are shown in Fig.\ref{f2}.a for a excitation fluence of $190mJ/cm^2$. Individually these dynamics contains all three degrees of freedom mentioned. However, by summing the effects under opposite biases we obtain the spin-averaged response, thus decoupling the charge and lattice degrees of freedom (shown in Fig.\ref{f2}.c) from spin dynamics. This charge-lattice combined response is characterised by a sharp transient, followed by a slow relaxation with a fast damping oscillations. Here the oscillations represent the coherent acoustic phonons and are independently extracted in Fig.\ref{f2}.d, thereby isolating the lattice dynamics. Similarly, the spin degree of freedom is singled out by subtracting the effects under opposite biases, as shown in Fig.\ref{f2}.b.

To further justify the significance of these additional dynamics, the constant bias experiments was performed again at excitation fluence of $48mJ/cm^2$ and $85mJ/cm^2$ as shown in Fig.\ref{f3}.a and Fig.\ref{f3}.c. Where the isolated spin and charge dynamics, corresponding to the magnetic and electronic responses are shown in Fig.\ref{f3}.b and Fig.\ref{f3}.d. It is shown that the magnetic response lags behind the electronic response for all values of fluence, and therefore substantiates the contribution of electronic and phononic dynamics in THz demagnetisation. Furthermore, it is shown that a sub picosecond magnetic response exists from a THz pulse excitation.

\section{\label{sec:level1}Conclusion}

In the present paper, the physical mechanisms behind nickel thin film demagnetisation due to femtosecond laser pulses are investigated. At low field amplitudes, we observe a linear response of demagnetisation from the intensity of the pump electromagnetic field. The THz pump fluence needed to achieve full demagnetisation is similar to the optical counterpart, and therefore suggesting that the frequency of the pump pulse does not play a major role in the demagnetisation process. To support this argument, it was found that at full demagnetisation, the THz pulse induces heating within the sample that is comparable to the Curie temperature of nickel. Alluding to sample heating as the main cause of the demagnetisation process.

\begin{acknowledgments}

We thank insightful discussions with Edoardo Baldini. 
CW was supported by grant EP/R02992X/1 from the UK Engineering and Physical Sciences Research Council (EPSRC).

% We wish to acknowledge the support of the author community in using
% REV\TeX{}, offering suggestions and encouragement, testing new versions,
% \dots.
\end{acknowledgments}

\bibliography{citations}% Produces the bibliography via BibTeX.

\providecommand{\noopsort}[1]{}\providecommand{\singleletter}[1]{#1}%
\begin{thebibliography}{30}%
\makeatletter
\providecommand \@ifxundefined [1]{%
 \@ifx{#1\undefined}
}%
\providecommand \@ifnum [1]{%
 \ifnum #1\expandafter \@firstoftwo
 \else \expandafter \@secondoftwo
 \fi
}%
\providecommand \@ifx [1]{%
 \ifx #1\expandafter \@firstoftwo
 \else \expandafter \@secondoftwo
 \fi
}%
\providecommand \natexlab [1]{#1}%
\providecommand \enquote  [1]{``#1''}%
\providecommand \bibnamefont  [1]{#1}%
\providecommand \bibfnamefont [1]{#1}%
\providecommand \citenamefont [1]{#1}%
\providecommand \href@noop [0]{\@secondoftwo}%
\providecommand \href [0]{\begingroup \@sanitize@url \@href}%
\providecommand \@href[1]{\@@startlink{#1}\@@href}%
\providecommand \@@href[1]{\endgroup#1\@@endlink}%
\providecommand \@sanitize@url [0]{\catcode `\\12\catcode `\$12\catcode
  `\&12\catcode `\#12\catcode `\^12\catcode `\_12\catcode `\%12\relax}%
\providecommand \@@startlink[1]{}%
\providecommand \@@endlink[0]{}%
\providecommand \url  [0]{\begingroup\@sanitize@url \@url }%
\providecommand \@url [1]{\endgroup\@href {#1}{\urlprefix }}%
\providecommand \urlprefix  [0]{URL }%
\providecommand \Eprint [0]{\href }%
\providecommand \doibase [0]{https://doi.org/}%
\providecommand \selectlanguage [0]{\@gobble}%
\providecommand \bibinfo  [0]{\@secondoftwo}%
\providecommand \bibfield  [0]{\@secondoftwo}%
\providecommand \translation [1]{[#1]}%
\providecommand \BibitemOpen [0]{}%
\providecommand \bibitemStop [0]{}%
\providecommand \bibitemNoStop [0]{.\EOS\space}%
\providecommand \EOS [0]{\spacefactor3000\relax}%
\providecommand \BibitemShut  [1]{\csname bibitem#1\endcsname}%
\let\auto@bib@innerbib\@empty
%</preamble>
\bibitem [{\citenamefont {Beaurepaire}\ \emph {et~al.}(1996)\citenamefont
  {Beaurepaire}, \citenamefont {Merle}, \citenamefont {Daunois},\ and\
  \citenamefont {Bigot}}]{manfred1}%
  \BibitemOpen
  \bibfield  {author} {\bibinfo {author} {\bibfnamefont {E.}~\bibnamefont
  {Beaurepaire}}, \bibinfo {author} {\bibfnamefont {J.-C.}\ \bibnamefont
  {Merle}}, \bibinfo {author} {\bibfnamefont {A.}~\bibnamefont {Daunois}},\
  and\ \bibinfo {author} {\bibfnamefont {J.-Y.}\ \bibnamefont {Bigot}},\
  }\bibfield  {title} {\bibinfo {title} {Ultrafast spin dynamics in
  ferromagnetic nickel},\ }\href {https://doi.org/10.1103/PhysRevLett.76.4250}
  {\bibfield  {journal} {\bibinfo  {journal} {Phys. Rev. Lett.}\ }\textbf
  {\bibinfo {volume} {76}},\ \bibinfo {pages} {4250} (\bibinfo {year}
  {1996})}\BibitemShut {NoStop}%
\bibitem [{\citenamefont {Shalaby}\ \emph {et~al.}(2016)\citenamefont
  {Shalaby}, \citenamefont {Vicario},\ and\ \citenamefont
  {Hauri}}]{mostafa_ni}%
  \BibitemOpen
  \bibfield  {author} {\bibinfo {author} {\bibfnamefont {M.}~\bibnamefont
  {Shalaby}}, \bibinfo {author} {\bibfnamefont {C.}~\bibnamefont {Vicario}},\
  and\ \bibinfo {author} {\bibfnamefont {C.~P.}\ \bibnamefont {Hauri}},\
  }\bibfield  {title} {\bibinfo {title} {Low frequency terahertz-induced
  demagnetization in ferromagnetic nickel},\ }\href
  {https://doi.org/10.1063/1.4948472} {\bibfield  {journal} {\bibinfo
  {journal} {Applied Physics Letters}\ }\textbf {\bibinfo {volume} {108}},\
  \bibinfo {pages} {182903} (\bibinfo {year} {2016})},\ \Eprint
  {https://arxiv.org/abs/https://doi.org/10.1063/1.4948472}
  {https://doi.org/10.1063/1.4948472} \BibitemShut {NoStop}%
\bibitem [{\citenamefont {Polley}\ \emph {et~al.}(2018)\citenamefont {Polley},
  \citenamefont {Pancaldi}, \citenamefont {Hudl}, \citenamefont {Vavassori},
  \citenamefont {Urazhdin},\ and\ \citenamefont {Bonetti}}]{thz}%
  \BibitemOpen
  \bibfield  {author} {\bibinfo {author} {\bibfnamefont {D.}~\bibnamefont
  {Polley}}, \bibinfo {author} {\bibfnamefont {M.}~\bibnamefont {Pancaldi}},
  \bibinfo {author} {\bibfnamefont {M.}~\bibnamefont {Hudl}}, \bibinfo {author}
  {\bibfnamefont {P.}~\bibnamefont {Vavassori}}, \bibinfo {author}
  {\bibfnamefont {S.}~\bibnamefont {Urazhdin}},\ and\ \bibinfo {author}
  {\bibfnamefont {S.}~\bibnamefont {Bonetti}},\ }\bibfield  {title} {\bibinfo
  {title} {{THz}-driven demagnetization with perpendicular magnetic anisotropy:
  towards ultrafast ballistic switching},\ }\href
  {https://doi.org/10.1088/1361-6463/aaa863} {\bibfield  {journal} {\bibinfo
  {journal} {Journal of Physics D: Applied Physics}\ }\textbf {\bibinfo
  {volume} {51}},\ \bibinfo {pages} {084001} (\bibinfo {year}
  {2018})}\BibitemShut {NoStop}%
\bibitem [{\citenamefont {{Hilton}}\ \emph {et~al.}(2005)\citenamefont
  {{Hilton}}, \citenamefont {{Averitt}}, \citenamefont {{Meserole}},
  \citenamefont {{Fisher}}, \citenamefont {{Funk}},\ and\ \citenamefont
  {{Taylor}}}]{thz2}%
  \BibitemOpen
  \bibfield  {author} {\bibinfo {author} {\bibfnamefont {D.~J.}\ \bibnamefont
  {{Hilton}}}, \bibinfo {author} {\bibfnamefont {R.~D.}\ \bibnamefont
  {{Averitt}}}, \bibinfo {author} {\bibfnamefont {C.~A.}\ \bibnamefont
  {{Meserole}}}, \bibinfo {author} {\bibfnamefont {G.~L.}\ \bibnamefont
  {{Fisher}}}, \bibinfo {author} {\bibfnamefont {D.~J.}\ \bibnamefont
  {{Funk}}},\ and\ \bibinfo {author} {\bibfnamefont {A.~J.}\ \bibnamefont
  {{Taylor}}},\ }\bibfield  {title} {\bibinfo {title} {Terahertz spectroscopy
  of ultrafast demagnetization in ferromagnetic iron},\ }in\ \href
  {https://doi.org/10.1109/QELS.2005.1548777} {\emph {\bibinfo {booktitle}
  {2005 Quantum Electronics and Laser Science Conference}}},\ Vol.~\bibinfo
  {volume} {1}\ (\bibinfo {year} {2005})\ pp.\ \bibinfo {pages} {347--349 Vol.
  1}\BibitemShut {NoStop}%
\bibitem [{\citenamefont {Hudl}\ \emph {et~al.}(2019)\citenamefont {Hudl},
  \citenamefont {d'Aquino}, \citenamefont {Pancaldi}, \citenamefont {Yang},
  \citenamefont {Samant}, \citenamefont {Parkin}, \citenamefont {D\"urr},
  \citenamefont {Serpico}, \citenamefont {Hoffmann},\ and\ \citenamefont
  {Bonetti}}]{referee1}%
  \BibitemOpen
  \bibfield  {author} {\bibinfo {author} {\bibfnamefont {M.}~\bibnamefont
  {Hudl}}, \bibinfo {author} {\bibfnamefont {M.}~\bibnamefont {d'Aquino}},
  \bibinfo {author} {\bibfnamefont {M.}~\bibnamefont {Pancaldi}}, \bibinfo
  {author} {\bibfnamefont {S.-H.}\ \bibnamefont {Yang}}, \bibinfo {author}
  {\bibfnamefont {M.~G.}\ \bibnamefont {Samant}}, \bibinfo {author}
  {\bibfnamefont {S.~S.~P.}\ \bibnamefont {Parkin}}, \bibinfo {author}
  {\bibfnamefont {H.~A.}\ \bibnamefont {D\"urr}}, \bibinfo {author}
  {\bibfnamefont {C.}~\bibnamefont {Serpico}}, \bibinfo {author} {\bibfnamefont
  {M.~C.}\ \bibnamefont {Hoffmann}},\ and\ \bibinfo {author} {\bibfnamefont
  {S.}~\bibnamefont {Bonetti}},\ }\bibfield  {title} {\bibinfo {title}
  {Nonlinear magnetization dynamics driven by strong terahertz fields},\ }\href
  {https://doi.org/10.1103/PhysRevLett.123.197204} {\bibfield  {journal}
  {\bibinfo  {journal} {Phys. Rev. Lett.}\ }\textbf {\bibinfo {volume} {123}},\
  \bibinfo {pages} {197204} (\bibinfo {year} {2019})}\BibitemShut {NoStop}%
\bibitem [{\citenamefont {Fähnle}\ \emph {et~al.}(2018)\citenamefont
  {Fähnle}, \citenamefont {Haag}, \citenamefont {Illg}, \citenamefont
  {Mueller}, \citenamefont {Weng}, \citenamefont {Tsatsoulis}, \citenamefont
  {Huang}, \citenamefont {Briones}, \citenamefont {Teeny}, \citenamefont
  {Zhang},\ and\ \citenamefont {Kuhn}}]{manfred2}%
  \BibitemOpen
  \bibfield  {author} {\bibinfo {author} {\bibfnamefont {M.}~\bibnamefont
  {Fähnle}}, \bibinfo {author} {\bibfnamefont {M.}~\bibnamefont {Haag}},
  \bibinfo {author} {\bibfnamefont {C.}~\bibnamefont {Illg}}, \bibinfo {author}
  {\bibfnamefont {B.}~\bibnamefont {Mueller}}, \bibinfo {author} {\bibfnamefont
  {W.}~\bibnamefont {Weng}}, \bibinfo {author} {\bibfnamefont {T.}~\bibnamefont
  {Tsatsoulis}}, \bibinfo {author} {\bibfnamefont {H.}~\bibnamefont {Huang}},
  \bibinfo {author} {\bibfnamefont {J.}~\bibnamefont {Briones}}, \bibinfo
  {author} {\bibfnamefont {N.}~\bibnamefont {Teeny}}, \bibinfo {author}
  {\bibfnamefont {L.}~\bibnamefont {Zhang}},\ and\ \bibinfo {author}
  {\bibfnamefont {T.}~\bibnamefont {Kuhn}},\ }\bibfield  {title} {\bibinfo
  {title} {Review of ultrafast demagnetization after femtosecond laser pulses:
  A complex interaction of light with quantum matter},\ }\href
  {https://doi.org/10.11648/j.ajmp.20180702.12} {\bibfield  {journal} {\bibinfo
   {journal} {Am. J. Mod. Phys.}\ }\textbf {\bibinfo {volume} {7}},\ \bibinfo
  {pages} {68} (\bibinfo {year} {2018})}\BibitemShut {NoStop}%
\bibitem [{\citenamefont {Koopmans}\ \emph {et~al.}(2010)\citenamefont
  {Koopmans}, \citenamefont {Malinowski}, \citenamefont {Dalla~Longa},
  \citenamefont {Steiauf}, \citenamefont {F{\"a}hnle}, \citenamefont {Roth},
  \citenamefont {Cinchetti},\ and\ \citenamefont
  {Aeschlimann}}]{electron-phonon}%
  \BibitemOpen
  \bibfield  {author} {\bibinfo {author} {\bibfnamefont {B.}~\bibnamefont
  {Koopmans}}, \bibinfo {author} {\bibfnamefont {G.}~\bibnamefont
  {Malinowski}}, \bibinfo {author} {\bibfnamefont {F.}~\bibnamefont
  {Dalla~Longa}}, \bibinfo {author} {\bibfnamefont {D.}~\bibnamefont
  {Steiauf}}, \bibinfo {author} {\bibfnamefont {M.}~\bibnamefont {F{\"a}hnle}},
  \bibinfo {author} {\bibfnamefont {T.}~\bibnamefont {Roth}}, \bibinfo {author}
  {\bibfnamefont {M.}~\bibnamefont {Cinchetti}},\ and\ \bibinfo {author}
  {\bibfnamefont {M.}~\bibnamefont {Aeschlimann}},\ }\bibfield  {title}
  {\bibinfo {title} {Explaining the paradoxical diversity of ultrafast
  laser-induced demagnetization},\ }\href {https://doi.org/10.1038/nmat2593}
  {\bibfield  {journal} {\bibinfo  {journal} {Nature Materials}\ }\textbf
  {\bibinfo {volume} {9}},\ \bibinfo {pages} {259} (\bibinfo {year}
  {2010})}\BibitemShut {NoStop}%
\bibitem [{\citenamefont {Kirilyuk}\ \emph {et~al.}(2013)\citenamefont
  {Kirilyuk}, \citenamefont {Kimel},\ and\ \citenamefont {Rasing}}]{optical}%
  \BibitemOpen
  \bibfield  {author} {\bibinfo {author} {\bibfnamefont {A.}~\bibnamefont
  {Kirilyuk}}, \bibinfo {author} {\bibfnamefont {A.~V.}\ \bibnamefont
  {Kimel}},\ and\ \bibinfo {author} {\bibfnamefont {T.}~\bibnamefont
  {Rasing}},\ }\bibfield  {title} {\bibinfo {title} {Laser-induced
  magnetization dynamics and reversal in ferrimagnetic alloys},\ }\href
  {https://doi.org/10.1088/0034-4885/76/2/026501} {\bibfield  {journal}
  {\bibinfo  {journal} {Reports on Progress in Physics}\ }\textbf {\bibinfo
  {volume} {76}},\ \bibinfo {pages} {026501} (\bibinfo {year}
  {2013})}\BibitemShut {NoStop}%
\bibitem [{\citenamefont {Bigot}\ \emph {et~al.}(2009)\citenamefont {Bigot},
  \citenamefont {Vomir},\ and\ \citenamefont {Beaurepaire}}]{manfred6}%
  \BibitemOpen
  \bibfield  {author} {\bibinfo {author} {\bibfnamefont {J.-Y.}\ \bibnamefont
  {Bigot}}, \bibinfo {author} {\bibfnamefont {M.}~\bibnamefont {Vomir}},\ and\
  \bibinfo {author} {\bibfnamefont {E.}~\bibnamefont {Beaurepaire}},\
  }\bibfield  {title} {\bibinfo {title} {Coherent ultrafast magnetism induced
  by femtosecond laser pulses},\ }\href {https://doi.org/10.1038/nphys1285}
  {\bibfield  {journal} {\bibinfo  {journal} {Nature Physics}\ }\textbf
  {\bibinfo {volume} {5}},\ \bibinfo {pages} {515} (\bibinfo {year}
  {2009})}\BibitemShut {NoStop}%
\bibitem [{\citenamefont {Saito}\ \emph {et~al.}(2003)\citenamefont {Saito},
  \citenamefont {Matsuda},\ and\ \citenamefont {Wright}}]{electron-phonon2}%
  \BibitemOpen
  \bibfield  {author} {\bibinfo {author} {\bibfnamefont {T.}~\bibnamefont
  {Saito}}, \bibinfo {author} {\bibfnamefont {O.}~\bibnamefont {Matsuda}},\
  and\ \bibinfo {author} {\bibfnamefont {O.~B.}\ \bibnamefont {Wright}},\
  }\bibfield  {title} {\bibinfo {title} {Picosecond acoustic phonon pulse
  generation in nickel and chromium},\ }\href
  {https://doi.org/10.1103/PhysRevB.67.205421} {\bibfield  {journal} {\bibinfo
  {journal} {Phys. Rev. B}\ }\textbf {\bibinfo {volume} {67}},\ \bibinfo
  {pages} {205421} (\bibinfo {year} {2003})}\BibitemShut {NoStop}%
\bibitem [{\citenamefont {van Kampen}\ \emph {et~al.}(2005)\citenamefont {van
  Kampen}, \citenamefont {Kohlhepp}, \citenamefont {de~Jonge}, \citenamefont
  {Koopmans},\ and\ \citenamefont {Coehoorn}}]{electron-phonon3}%
  \BibitemOpen
  \bibfield  {author} {\bibinfo {author} {\bibfnamefont {M.}~\bibnamefont {van
  Kampen}}, \bibinfo {author} {\bibfnamefont {J.~T.}\ \bibnamefont {Kohlhepp}},
  \bibinfo {author} {\bibfnamefont {W.~J.~M.}\ \bibnamefont {de~Jonge}},
  \bibinfo {author} {\bibfnamefont {B.}~\bibnamefont {Koopmans}},\ and\
  \bibinfo {author} {\bibfnamefont {R.}~\bibnamefont {Coehoorn}},\ }\bibfield
  {title} {\bibinfo {title} {Sub-picosecond electron and phonon dynamics in
  nickel},\ }\href {https://doi.org/10.1088/0953-8984/17/43/004} {\bibfield
  {journal} {\bibinfo  {journal} {Journal of Physics: Condensed Matter}\
  }\textbf {\bibinfo {volume} {17}},\ \bibinfo {pages} {6823} (\bibinfo {year}
  {2005})}\BibitemShut {NoStop}%
\bibitem [{\citenamefont {Bonetti}\ \emph {et~al.}(2016)\citenamefont
  {Bonetti}, \citenamefont {Hoffmann}, \citenamefont {Sher}, \citenamefont
  {Chen}, \citenamefont {Yang}, \citenamefont {Samant}, \citenamefont
  {Parkin},\ and\ \citenamefont {D\"urr}}]{electron-phonon4}%
  \BibitemOpen
  \bibfield  {author} {\bibinfo {author} {\bibfnamefont {S.}~\bibnamefont
  {Bonetti}}, \bibinfo {author} {\bibfnamefont {M.~C.}\ \bibnamefont
  {Hoffmann}}, \bibinfo {author} {\bibfnamefont {M.-J.}\ \bibnamefont {Sher}},
  \bibinfo {author} {\bibfnamefont {Z.}~\bibnamefont {Chen}}, \bibinfo {author}
  {\bibfnamefont {S.-H.}\ \bibnamefont {Yang}}, \bibinfo {author}
  {\bibfnamefont {M.~G.}\ \bibnamefont {Samant}}, \bibinfo {author}
  {\bibfnamefont {S.~S.~P.}\ \bibnamefont {Parkin}},\ and\ \bibinfo {author}
  {\bibfnamefont {H.~A.}\ \bibnamefont {D\"urr}},\ }\bibfield  {title}
  {\bibinfo {title} {Thz-driven ultrafast spin-lattice scattering in amorphous
  metallic ferromagnets},\ }\href
  {https://doi.org/10.1103/PhysRevLett.117.087205} {\bibfield  {journal}
  {\bibinfo  {journal} {Phys. Rev. Lett.}\ }\textbf {\bibinfo {volume} {117}},\
  \bibinfo {pages} {087205} (\bibinfo {year} {2016})}\BibitemShut {NoStop}%
\bibitem [{\citenamefont {Zhang}\ \emph {et~al.}(2009)\citenamefont {Zhang},
  \citenamefont {H{\"u}bner}, \citenamefont {Lefkidis}, \citenamefont {Bai},\
  and\ \citenamefont {George}}]{manfred5}%
  \BibitemOpen
  \bibfield  {author} {\bibinfo {author} {\bibfnamefont {G.~P.}\ \bibnamefont
  {Zhang}}, \bibinfo {author} {\bibfnamefont {W.}~\bibnamefont {H{\"u}bner}},
  \bibinfo {author} {\bibfnamefont {G.}~\bibnamefont {Lefkidis}}, \bibinfo
  {author} {\bibfnamefont {Y.}~\bibnamefont {Bai}},\ and\ \bibinfo {author}
  {\bibfnamefont {T.~F.}\ \bibnamefont {George}},\ }\bibfield  {title}
  {\bibinfo {title} {Paradigm of the time-resolved magneto-optical kerr effect
  for femtosecond magnetism},\ }\href {https://doi.org/10.1038/nphys1315}
  {\bibfield  {journal} {\bibinfo  {journal} {Nature Physics}\ }\textbf
  {\bibinfo {volume} {5}},\ \bibinfo {pages} {499} (\bibinfo {year}
  {2009})}\BibitemShut {NoStop}%
\bibitem [{\citenamefont {Mueller}\ \emph {et~al.}(2013)\citenamefont
  {Mueller}, \citenamefont {Baral}, \citenamefont {Vollmar}, \citenamefont
  {Cinchetti}, \citenamefont {Aeschlimann}, \citenamefont {Schneider},\ and\
  \citenamefont {Rethfeld}}]{spin-flip}%
  \BibitemOpen
  \bibfield  {author} {\bibinfo {author} {\bibfnamefont {B.~Y.}\ \bibnamefont
  {Mueller}}, \bibinfo {author} {\bibfnamefont {A.}~\bibnamefont {Baral}},
  \bibinfo {author} {\bibfnamefont {S.}~\bibnamefont {Vollmar}}, \bibinfo
  {author} {\bibfnamefont {M.}~\bibnamefont {Cinchetti}}, \bibinfo {author}
  {\bibfnamefont {M.}~\bibnamefont {Aeschlimann}}, \bibinfo {author}
  {\bibfnamefont {H.~C.}\ \bibnamefont {Schneider}},\ and\ \bibinfo {author}
  {\bibfnamefont {B.}~\bibnamefont {Rethfeld}},\ }\bibfield  {title} {\bibinfo
  {title} {Feedback effect during ultrafast demagnetization dynamics in
  ferromagnets},\ }\href {https://doi.org/10.1103/PhysRevLett.111.167204}
  {\bibfield  {journal} {\bibinfo  {journal} {Phys. Rev. Lett.}\ }\textbf
  {\bibinfo {volume} {111}},\ \bibinfo {pages} {167204} (\bibinfo {year}
  {2013})}\BibitemShut {NoStop}%
\bibitem [{\citenamefont {Stamm}\ \emph {et~al.}(2007)\citenamefont {Stamm},
  \citenamefont {Kachel}, \citenamefont {Pontius}, \citenamefont {Mitzner},
  \citenamefont {Quast}, \citenamefont {Holldack}, \citenamefont {Khan},
  \citenamefont {Lupulescu}, \citenamefont {Aziz}, \citenamefont {Wietstruk},
  \citenamefont {D{\"u}rr},\ and\ \citenamefont {Eberhardt}}]{spin-flip2}%
  \BibitemOpen
  \bibfield  {author} {\bibinfo {author} {\bibfnamefont {C.}~\bibnamefont
  {Stamm}}, \bibinfo {author} {\bibfnamefont {T.}~\bibnamefont {Kachel}},
  \bibinfo {author} {\bibfnamefont {N.}~\bibnamefont {Pontius}}, \bibinfo
  {author} {\bibfnamefont {R.}~\bibnamefont {Mitzner}}, \bibinfo {author}
  {\bibfnamefont {T.}~\bibnamefont {Quast}}, \bibinfo {author} {\bibfnamefont
  {K.}~\bibnamefont {Holldack}}, \bibinfo {author} {\bibfnamefont
  {S.}~\bibnamefont {Khan}}, \bibinfo {author} {\bibfnamefont {C.}~\bibnamefont
  {Lupulescu}}, \bibinfo {author} {\bibfnamefont {E.~F.}\ \bibnamefont {Aziz}},
  \bibinfo {author} {\bibfnamefont {M.}~\bibnamefont {Wietstruk}}, \bibinfo
  {author} {\bibfnamefont {H.~A.}\ \bibnamefont {D{\"u}rr}},\ and\ \bibinfo
  {author} {\bibfnamefont {W.}~\bibnamefont {Eberhardt}},\ }\bibfield  {title}
  {\bibinfo {title} {Femtosecond modification of electron localization and
  transfer of angular momentum in nickel},\ }\href
  {https://doi.org/10.1038/nmat1985} {\bibfield  {journal} {\bibinfo  {journal}
  {Nature Materials}\ }\textbf {\bibinfo {volume} {6}},\ \bibinfo {pages} {740}
  (\bibinfo {year} {2007})}\BibitemShut {NoStop}%
\bibitem [{\citenamefont {Cinchetti}\ \emph {et~al.}(2006)\citenamefont
  {Cinchetti}, \citenamefont {S\'anchez~Albaneda}, \citenamefont {Hoffmann},
  \citenamefont {Roth}, \citenamefont {W\"ustenberg}, \citenamefont
  {Krau\ss{}}, \citenamefont {Andreyev}, \citenamefont {Schneider},
  \citenamefont {Bauer},\ and\ \citenamefont {Aeschlimann}}]{spin-flip3}%
  \BibitemOpen
  \bibfield  {author} {\bibinfo {author} {\bibfnamefont {M.}~\bibnamefont
  {Cinchetti}}, \bibinfo {author} {\bibfnamefont {M.}~\bibnamefont
  {S\'anchez~Albaneda}}, \bibinfo {author} {\bibfnamefont {D.}~\bibnamefont
  {Hoffmann}}, \bibinfo {author} {\bibfnamefont {T.}~\bibnamefont {Roth}},
  \bibinfo {author} {\bibfnamefont {J.-P.}\ \bibnamefont {W\"ustenberg}},
  \bibinfo {author} {\bibfnamefont {M.}~\bibnamefont {Krau\ss{}}}, \bibinfo
  {author} {\bibfnamefont {O.}~\bibnamefont {Andreyev}}, \bibinfo {author}
  {\bibfnamefont {H.~C.}\ \bibnamefont {Schneider}}, \bibinfo {author}
  {\bibfnamefont {M.}~\bibnamefont {Bauer}},\ and\ \bibinfo {author}
  {\bibfnamefont {M.}~\bibnamefont {Aeschlimann}},\ }\bibfield  {title}
  {\bibinfo {title} {Spin-flip processes and ultrafast magnetization dynamics
  in co: Unifying the microscopic and macroscopic view of femtosecond
  magnetism},\ }\href {https://doi.org/10.1103/PhysRevLett.97.177201}
  {\bibfield  {journal} {\bibinfo  {journal} {Phys. Rev. Lett.}\ }\textbf
  {\bibinfo {volume} {97}},\ \bibinfo {pages} {177201} (\bibinfo {year}
  {2006})}\BibitemShut {NoStop}%
\bibitem [{\citenamefont {Carpene}\ \emph {et~al.}(2008)\citenamefont
  {Carpene}, \citenamefont {Mancini}, \citenamefont {Dallera}, \citenamefont
  {Brenna}, \citenamefont {Puppin},\ and\ \citenamefont
  {De~Silvestri}}]{magnon}%
  \BibitemOpen
  \bibfield  {author} {\bibinfo {author} {\bibfnamefont {E.}~\bibnamefont
  {Carpene}}, \bibinfo {author} {\bibfnamefont {E.}~\bibnamefont {Mancini}},
  \bibinfo {author} {\bibfnamefont {C.}~\bibnamefont {Dallera}}, \bibinfo
  {author} {\bibfnamefont {M.}~\bibnamefont {Brenna}}, \bibinfo {author}
  {\bibfnamefont {E.}~\bibnamefont {Puppin}},\ and\ \bibinfo {author}
  {\bibfnamefont {S.}~\bibnamefont {De~Silvestri}},\ }\bibfield  {title}
  {\bibinfo {title} {Dynamics of electron-magnon interaction and ultrafast
  demagnetization in thin iron films},\ }\href
  {https://doi.org/10.1103/PhysRevB.78.174422} {\bibfield  {journal} {\bibinfo
  {journal} {Phys. Rev. B}\ }\textbf {\bibinfo {volume} {78}},\ \bibinfo
  {pages} {174422} (\bibinfo {year} {2008})}\BibitemShut {NoStop}%
\bibitem [{\citenamefont {Mueller}\ \emph {et~al.}(2011)\citenamefont
  {Mueller}, \citenamefont {Roth}, \citenamefont {Cinchetti}, \citenamefont
  {Aeschlimann},\ and\ \citenamefont {Rethfeld}}]{manfred4}%
  \BibitemOpen
  \bibfield  {author} {\bibinfo {author} {\bibfnamefont {B.~Y.}\ \bibnamefont
  {Mueller}}, \bibinfo {author} {\bibfnamefont {T.}~\bibnamefont {Roth}},
  \bibinfo {author} {\bibfnamefont {M.}~\bibnamefont {Cinchetti}}, \bibinfo
  {author} {\bibfnamefont {M.}~\bibnamefont {Aeschlimann}},\ and\ \bibinfo
  {author} {\bibfnamefont {B.}~\bibnamefont {Rethfeld}},\ }\bibfield  {title}
  {\bibinfo {title} {Driving force of ultrafast magnetization dynamics},\
  }\href {https://doi.org/10.1088/1367-2630/13/12/123010} {\bibfield  {journal}
  {\bibinfo  {journal} {New Journal of Physics}\ }\textbf {\bibinfo {volume}
  {13}},\ \bibinfo {pages} {123010} (\bibinfo {year} {2011})}\BibitemShut
  {NoStop}%
\bibitem [{\citenamefont {Haag}\ \emph {et~al.}(2014)\citenamefont {Haag},
  \citenamefont {Illg},\ and\ \citenamefont {F\"ahnle}}]{chem_pot}%
  \BibitemOpen
  \bibfield  {author} {\bibinfo {author} {\bibfnamefont {M.}~\bibnamefont
  {Haag}}, \bibinfo {author} {\bibfnamefont {C.}~\bibnamefont {Illg}},\ and\
  \bibinfo {author} {\bibfnamefont {M.}~\bibnamefont {F\"ahnle}},\ }\bibfield
  {title} {\bibinfo {title} {Role of electron-magnon scatterings in ultrafast
  demagnetization},\ }\href {https://doi.org/10.1103/PhysRevB.90.014417}
  {\bibfield  {journal} {\bibinfo  {journal} {Phys. Rev. B}\ }\textbf {\bibinfo
  {volume} {90}},\ \bibinfo {pages} {014417} (\bibinfo {year}
  {2014})}\BibitemShut {NoStop}%
\bibitem [{\citenamefont {Battiato}\ \emph {et~al.}(2010)\citenamefont
  {Battiato}, \citenamefont {Carva},\ and\ \citenamefont
  {Oppeneer}}]{manfred3}%
  \BibitemOpen
  \bibfield  {author} {\bibinfo {author} {\bibfnamefont {M.}~\bibnamefont
  {Battiato}}, \bibinfo {author} {\bibfnamefont {K.}~\bibnamefont {Carva}},\
  and\ \bibinfo {author} {\bibfnamefont {P.~M.}\ \bibnamefont {Oppeneer}},\
  }\bibfield  {title} {\bibinfo {title} {Superdiffusive spin transport as a
  mechanism of ultrafast demagnetization},\ }\href
  {https://doi.org/10.1103/PhysRevLett.105.027203} {\bibfield  {journal}
  {\bibinfo  {journal} {Phys. Rev. Lett.}\ }\textbf {\bibinfo {volume} {105}},\
  \bibinfo {pages} {027203} (\bibinfo {year} {2010})}\BibitemShut {NoStop}%
\bibitem [{\citenamefont {Wang}\ \emph {et~al.}(2018)\citenamefont {Wang},
  \citenamefont {Kovalev}, \citenamefont {Awari}, \citenamefont {Chen},
  \citenamefont {Germanskiy}, \citenamefont {Green}, \citenamefont {Deinert},
  \citenamefont {Kampfrath}, \citenamefont {Milano},\ and\ \citenamefont
  {Gensch}}]{kampfrath}%
  \BibitemOpen
  \bibfield  {author} {\bibinfo {author} {\bibfnamefont {Z.}~\bibnamefont
  {Wang}}, \bibinfo {author} {\bibfnamefont {S.}~\bibnamefont {Kovalev}},
  \bibinfo {author} {\bibfnamefont {N.}~\bibnamefont {Awari}}, \bibinfo
  {author} {\bibfnamefont {M.}~\bibnamefont {Chen}}, \bibinfo {author}
  {\bibfnamefont {S.}~\bibnamefont {Germanskiy}}, \bibinfo {author}
  {\bibfnamefont {B.}~\bibnamefont {Green}}, \bibinfo {author} {\bibfnamefont
  {J.-C.}\ \bibnamefont {Deinert}}, \bibinfo {author} {\bibfnamefont
  {T.}~\bibnamefont {Kampfrath}}, \bibinfo {author} {\bibfnamefont
  {J.}~\bibnamefont {Milano}},\ and\ \bibinfo {author} {\bibfnamefont
  {M.}~\bibnamefont {Gensch}},\ }\bibfield  {title} {\bibinfo {title} {Magnetic
  field dependence of antiferromagnetic resonance in nio},\ }\href
  {https://doi.org/10.1063/1.5031213} {\bibfield  {journal} {\bibinfo
  {journal} {Applied Physics Letters}\ }\textbf {\bibinfo {volume} {112}},\
  \bibinfo {pages} {252404} (\bibinfo {year} {2018})},\ \Eprint
  {https://arxiv.org/abs/https://doi.org/10.1063/1.5031213}
  {https://doi.org/10.1063/1.5031213} \BibitemShut {NoStop}%
\bibitem [{\citenamefont {Roth}\ \emph {et~al.}(2012)\citenamefont {Roth},
  \citenamefont {Schellekens}, \citenamefont {Alebrand}, \citenamefont
  {Schmitt}, \citenamefont {Steil}, \citenamefont {Koopmans}, \citenamefont
  {Cinchetti},\ and\ \citenamefont {Aeschlimann}}]{temp}%
  \BibitemOpen
  \bibfield  {author} {\bibinfo {author} {\bibfnamefont {T.}~\bibnamefont
  {Roth}}, \bibinfo {author} {\bibfnamefont {A.~J.}\ \bibnamefont
  {Schellekens}}, \bibinfo {author} {\bibfnamefont {S.}~\bibnamefont
  {Alebrand}}, \bibinfo {author} {\bibfnamefont {O.}~\bibnamefont {Schmitt}},
  \bibinfo {author} {\bibfnamefont {D.}~\bibnamefont {Steil}}, \bibinfo
  {author} {\bibfnamefont {B.}~\bibnamefont {Koopmans}}, \bibinfo {author}
  {\bibfnamefont {M.}~\bibnamefont {Cinchetti}},\ and\ \bibinfo {author}
  {\bibfnamefont {M.}~\bibnamefont {Aeschlimann}},\ }\bibfield  {title}
  {\bibinfo {title} {Temperature dependence of laser-induced demagnetization in
  ni: A key for identifying the underlying mechanism},\ }\href
  {https://doi.org/10.1103/PhysRevX.2.021006} {\bibfield  {journal} {\bibinfo
  {journal} {Phys. Rev. X}\ }\textbf {\bibinfo {volume} {2}},\ \bibinfo {pages}
  {021006} (\bibinfo {year} {2012})}\BibitemShut {NoStop}%
\bibitem [{\citenamefont {Atxitia}\ \emph {et~al.}(2010)\citenamefont
  {Atxitia}, \citenamefont {Chubykalo-Fesenko}, \citenamefont {Walowski},
  \citenamefont {Mann},\ and\ \citenamefont {M\"unzenberg}}]{temp2}%
  \BibitemOpen
  \bibfield  {author} {\bibinfo {author} {\bibfnamefont {U.}~\bibnamefont
  {Atxitia}}, \bibinfo {author} {\bibfnamefont {O.}~\bibnamefont
  {Chubykalo-Fesenko}}, \bibinfo {author} {\bibfnamefont {J.}~\bibnamefont
  {Walowski}}, \bibinfo {author} {\bibfnamefont {A.}~\bibnamefont {Mann}},\
  and\ \bibinfo {author} {\bibfnamefont {M.}~\bibnamefont {M\"unzenberg}},\
  }\bibfield  {title} {\bibinfo {title} {Evidence for thermal mechanisms in
  laser-induced femtosecond spin dynamics},\ }\href
  {https://doi.org/10.1103/PhysRevB.81.174401} {\bibfield  {journal} {\bibinfo
  {journal} {Phys. Rev. B}\ }\textbf {\bibinfo {volume} {81}},\ \bibinfo
  {pages} {174401} (\bibinfo {year} {2010})}\BibitemShut {NoStop}%
\bibitem [{\citenamefont {Steiauf}\ \emph {et~al.}(2010)\citenamefont
  {Steiauf}, \citenamefont {Illg},\ and\ \citenamefont {Fähnle}}]{ey}%
  \BibitemOpen
  \bibfield  {author} {\bibinfo {author} {\bibfnamefont {D.}~\bibnamefont
  {Steiauf}}, \bibinfo {author} {\bibfnamefont {C.}~\bibnamefont {Illg}},\ and\
  \bibinfo {author} {\bibfnamefont {M.}~\bibnamefont {Fähnle}},\ }\bibfield
  {title} {\bibinfo {title} {Demagnetization on the fs time-scale by the
  elliott-yafet mechanism},\ }\href
  {https://doi.org/10.1088/1742-6596/200/4/042024} {\bibfield  {journal}
  {\bibinfo  {journal} {Journal of Physics: Conference Series}\ }\textbf
  {\bibinfo {volume} {200}},\ \bibinfo {pages} {042024} (\bibinfo {year}
  {2010})}\BibitemShut {NoStop}%
\bibitem [{\citenamefont {Vicario}\ \emph {et~al.}(2013)\citenamefont
  {Vicario}, \citenamefont {Ruchert}, \citenamefont {Ardana-Lamas},
  \citenamefont {Derlet}, \citenamefont {Tudu}, \citenamefont {Luning},\ and\
  \citenamefont {Hauri}}]{cobalt}%
  \BibitemOpen
  \bibfield  {author} {\bibinfo {author} {\bibfnamefont {C.}~\bibnamefont
  {Vicario}}, \bibinfo {author} {\bibfnamefont {C.}~\bibnamefont {Ruchert}},
  \bibinfo {author} {\bibfnamefont {F.}~\bibnamefont {Ardana-Lamas}}, \bibinfo
  {author} {\bibfnamefont {P.~M.}\ \bibnamefont {Derlet}}, \bibinfo {author}
  {\bibfnamefont {B.}~\bibnamefont {Tudu}}, \bibinfo {author} {\bibfnamefont
  {J.}~\bibnamefont {Luning}},\ and\ \bibinfo {author} {\bibfnamefont {C.~P.}\
  \bibnamefont {Hauri}},\ }\bibfield  {title} {\bibinfo {title} {Off-resonant
  magnetization dynamics phase-locked to an intense phase-stable terahertz
  transient},\ }\href {https://doi.org/10.1038/nphoton.2013.209} {\bibfield
  {journal} {\bibinfo  {journal} {Nature Photonics}\ }\textbf {\bibinfo
  {volume} {7}},\ \bibinfo {pages} {720} (\bibinfo {year} {2013})}\BibitemShut
  {NoStop}%
\bibitem [{\citenamefont {Shalaby}\ \emph {et~al.}(2018)\citenamefont
  {Shalaby}, \citenamefont {Donges}, \citenamefont {Carva}, \citenamefont
  {Allenspach}, \citenamefont {Oppeneer}, \citenamefont {Nowak},\ and\
  \citenamefont {Hauri}}]{mokekerr}%
  \BibitemOpen
  \bibfield  {author} {\bibinfo {author} {\bibfnamefont {M.}~\bibnamefont
  {Shalaby}}, \bibinfo {author} {\bibfnamefont {A.}~\bibnamefont {Donges}},
  \bibinfo {author} {\bibfnamefont {K.}~\bibnamefont {Carva}}, \bibinfo
  {author} {\bibfnamefont {R.}~\bibnamefont {Allenspach}}, \bibinfo {author}
  {\bibfnamefont {P.~M.}\ \bibnamefont {Oppeneer}}, \bibinfo {author}
  {\bibfnamefont {U.}~\bibnamefont {Nowak}},\ and\ \bibinfo {author}
  {\bibfnamefont {C.~P.}\ \bibnamefont {Hauri}},\ }\bibfield  {title} {\bibinfo
  {title} {Coherent and incoherent ultrafast magnetization dynamics in $3d$
  ferromagnets driven by extreme terahertz fields},\ }\href
  {https://doi.org/10.1103/PhysRevB.98.014405} {\bibfield  {journal} {\bibinfo
  {journal} {Phys. Rev. B}\ }\textbf {\bibinfo {volume} {98}},\ \bibinfo
  {pages} {014405} (\bibinfo {year} {2018})}\BibitemShut {NoStop}%
\bibitem [{\citenamefont {Shalaby}\ \emph {et~al.}(2017)\citenamefont
  {Shalaby}, \citenamefont {Vicario},\ and\ \citenamefont {Hauri}}]{diamond}%
  \BibitemOpen
  \bibfield  {author} {\bibinfo {author} {\bibfnamefont {M.}~\bibnamefont
  {Shalaby}}, \bibinfo {author} {\bibfnamefont {C.}~\bibnamefont {Vicario}},\
  and\ \bibinfo {author} {\bibfnamefont {C.~P.}\ \bibnamefont {Hauri}},\
  }\bibfield  {title} {\bibinfo {title} {Extreme nonlinear terahertz
  electro-optics in diamond for ultrafast pulse switching},\ }\href
  {https://doi.org/10.1063/1.4978051} {\bibfield  {journal} {\bibinfo
  {journal} {APL Photonics}\ }\textbf {\bibinfo {volume} {2}},\ \bibinfo
  {pages} {036106} (\bibinfo {year} {2017})},\ \Eprint
  {https://arxiv.org/abs/https://doi.org/10.1063/1.4978051}
  {https://doi.org/10.1063/1.4978051} \BibitemShut {NoStop}%
\bibitem [{\citenamefont {Shalaby}\ and\ \citenamefont
  {Hauri}(2015)}]{thz_bullet}%
  \BibitemOpen
  \bibfield  {author} {\bibinfo {author} {\bibfnamefont {M.}~\bibnamefont
  {Shalaby}}\ and\ \bibinfo {author} {\bibfnamefont {C.~P.}\ \bibnamefont
  {Hauri}},\ }\bibfield  {title} {\bibinfo {title} {Demonstration of a
  low-frequency three-dimensional terahertz bullet with extreme brightness},\
  }\href {https://doi.org/10.1038/ncomms6976} {\bibfield  {journal} {\bibinfo
  {journal} {Nature Communications}\ }\textbf {\bibinfo {volume} {6}},\
  \bibinfo {pages} {5976} (\bibinfo {year} {2015})}\BibitemShut {NoStop}%
\bibitem [{\citenamefont {Hwang}\ and\ \citenamefont
  {Das~Sarma}(2013)}]{hot_e1}%
  \BibitemOpen
  \bibfield  {author} {\bibinfo {author} {\bibfnamefont {E.~H.}\ \bibnamefont
  {Hwang}}\ and\ \bibinfo {author} {\bibfnamefont {S.}~\bibnamefont
  {Das~Sarma}},\ }\bibfield  {title} {\bibinfo {title} {Surface polar optical
  phonon interaction induced many-body effects and hot-electron relaxation in
  graphene},\ }\href {https://doi.org/10.1103/PhysRevB.87.115432} {\bibfield
  {journal} {\bibinfo  {journal} {Phys. Rev. B}\ }\textbf {\bibinfo {volume}
  {87}},\ \bibinfo {pages} {115432} (\bibinfo {year} {2013})}\BibitemShut
  {NoStop}%
\bibitem [{\citenamefont {Kalafati}\ and\ \citenamefont
  {Kokin}(1992)}]{hot_e2}%
  \BibitemOpen
  \bibfield  {author} {\bibinfo {author} {\bibfnamefont {Y.~D.}\ \bibnamefont
  {Kalafati}}\ and\ \bibinfo {author} {\bibfnamefont {V.~A.}\ \bibnamefont
  {Kokin}},\ }\bibfield  {title} {\bibinfo {title} {Hot carrier effects on
  nonlinear picosecond dynamics of semiconductor lasers and superluminescent
  diodes},\ }in\ \href
  {http://www.osapublishing.org/abstract.cfm?URI=QELS-1992-QThD15} {\emph
  {\bibinfo {booktitle} {Quantum Electronics and Laser Science Conference}}}\
  (\bibinfo  {publisher} {Optical Society of America},\ \bibinfo {year}
  {1992})\ p.\ \bibinfo {pages} {QThD15}\BibitemShut {NoStop}%
\end{thebibliography}%

\end{document}